\documentclass[manuscript]{aastex}

\begin{document}


\title{Near-UV Sources in the Hubble Ultra Deep Field: The Catalog \\
    }


\author{Elysse N. Voyer\altaffilmark{1,2}, Duilia F. de Mello\altaffilmark{1,3}}
\affil{Catholic University of America Washington, DC 20064}

\author{Brian Siana}
\affil{California Institute of Technology, MS 105-24, Pasadena, CA 91125}

\author{Jonathan P. Gardner}
\affil{Observational Cosmology Laboratory, Goddard Space Flight Center, Code 665, Greenbelt, MD 20771}

\author{Cori Quirk} 
\affil{Catholic University of America Washington, DC 20064}

\and

\author{Harry I. Teplitz}
\affil{Spitzer Science Center, California Institute of Technology, MS 220-6, Pasadena, CA 91125}


\altaffiltext{1}{Observational Cosmology Laboratory, Goddard Space Flight Center, Code 665, Greenbelt, MD 20771}
\altaffiltext{2}{NASA Graduate Student Research Program Fellow}
\altaffiltext{3}{Visiting Scientist in the Department of Physics and Astronomy, Johns Hopkins University, Baltimore, MD 21218}


\begin{abstract}

The catalog from the first high resolution U-band image of the Hubble Ultra Deep Field, taken with Hubble's Wide Field Planetary Camera 2 through the 
F300W filter, is presented. We detect 96 U-band objects and compare and combine this catalog with
a Great Observatories Origins Deep Survey (GOODS) B-selected catalog that provides B, V, i, and z photometry, spectral types, and photometric redshifts. 
We have also obtained Far-Ultraviolet (FUV, 1614 \AA) data with Hubble's Advanced Camera for Surveys Solar Blind Channel (ACS/SBC) and with Galaxy Evolution Explorer (GALEX). We detected 
31 sources with ACS/SBC, 28 with GALEX/FUV, and 45 with GALEX/NUV. The methods of 
observations, image processing, object identification, catalog preparation, and catalog matching are presented.

\end{abstract}






\section{Introduction}

The Hubble Ultra Deep Field (UDF) campaign (Beckwith et al. 2006) has produced the 
deepest optical images of our universe to date.  
The UDF was observed by Hubble in 412 orbits that were centered in a region of the 
Chandra Deep Field South (CDF-S) which was also the target of 
the Great Observatories Origins Deep Survey (GOODS, Giavalisco et al.
2004) known as the GOODS south or GOODS-S. The UDF used the same Advanced Camera for Surveys (ACS) filters as GOODS, 
F435W (B$_{\rm 435}$), F606W (V$_{\rm 606}$), F775W (i$_{\rm 775}$), and F850LP (z$_{\rm 850}$), but 
covered only one field of the 15 GOODS-S fields. 
The UDF reached approximately uniform limiting 
magnitudes m$_{\rm AB}$$\sim$ 29 for point sources, at least two magnitudes deeper than GOODS.
Both campaigns, the UDF and GOODS, did not include deep imaging in the U bandpass. 
Taking U-band photometry is a time 
consuming task because longer integrations are required to achieve comparable depth to optical images. 
Only three HST U-band deep fields have been taken so far, the original Hubble
Deep Field, the Hubble Deep Field South and the deepest U-band ever taken with Hubble which was 
part of the parallel campaign of the UDF and lies on the edge of the GOODS-S
(Williams et al. 1996, Casertano et al. 2000, de Mello et al. 2006b). GOODS 
has only partial U-band coverage with HST obtained during parallel observations (de Mello et al. 2006a). 
Deep U-band ground-based images of the GOODS-S field, such as those taken with the CTIO 4m and ESO 2.2m available in the GOODS
webpage\footnote[5]{http://www.stsci.edu/science/goods/SupportObs/cdfs$_{-}$mosaic/}, are included in 
the multiwavelength coverage of GOODS. Although ground-based observations can cover larger fields of view
than Hubble's cameras more efficiently, they do not possess the same angular resolution as space-based observations. 
Low-resolution ground-based images will blend together nearby detections 
leading to inaccurate photometric redshifts and morphological analysis. 
The U-band is a critical wavelength in studies 
at intermediate redshifts (z $<$ 2) since the rest-frame UV light is redshifted into the U bandpass. 
It is in the UV that short-lived, massive, O and B stars radiate most of their energy and therefore 
the U-band is necessary to probe the unobscured star-formation activity in galaxies at z $<$ 2. 

After a redshift of z$\sim$1.5-2, the star-formation rate of the universe began to steadily decline, decreasing by more than an order of magnitude (e.g. Hopkins \& Beacom 2006, 
Wadadekar, Casertano \& de Mello 2006). 
However, it is still an open question as to what population
of objects contributes to the SFR density during the decline, and whether downsizing (the shift in star-formation being dominated from
large to small mass galaxies as the universe aged) plays an important role in this era (Cowie et al. 1996, Savaglio et al. 2005, Mouri and Taniguchi 2006, Neistein, van den Bosch, \& Dekel 2006).     
Therefore, studies that use U-band observations to better understand the nature of star-forming galaxies at intermediate-z can greatly contribute to connecting 
the early universe and the local universe.    

In this paper we present the first HST targeted U-band image of the UDF. 
We present the U-band object catalog, and describe the methods used for observations, image processing, object 
identification, and catalog preparation.

\section{Observations}

The U-band observations were obtained with the HST Wide Field Planetary Camera 2 (WFPC2) 
in the Cycle 13 HST Treasury proposal (Teplitz, Program 10403). The UDF was imaged in the near UV (NUV) 
through the WFPC2/F300W filter ($\lambda$$_{\rm max}$ = 2987\AA, $\Delta$$\lambda$=740\AA) in 12 HST orbits divided into 4 roll angles to compensate for the 
shape of the WFPC2 chevron and achieve uniform depth. On-chip binning (2x2) was applied during WFPC2 observations to reduce the effects 
of read-out noise, i.e. each Wide Field became 400 $\times$ 400 pixels. A total of 24 WFPC2 images were taken with individual exposure times of 1200s. 

FUV imaging was also obtained in this proposal with HST/ACS Solar Blind 
Channel (SBC) camera at which time WFPC2 parallel observations were also made. 
The UDF ACS/SBC images were taken in 50 HST orbits using the long-pass
quartz filter (F150LP) ($\lambda$$_{\rm eff}$ = 1614\AA, and FWHM=177\AA). The FUV was imaged in 25 pointings. Each pointing had a four point dither pattern with two 
650s exposures at each dither position. The total exposure time per pointing was $\sim$5200s (Siana et al. 2007).

\section{Image Processing}

The WFPC2 images were retrieved from the HST archive for further processing. We combined the 24 WFPC2 images 
with the MultiDrizzle code in the PyDrizzle package (Koekemoer et al. 2002). These images were taken using a dithering technique that reduces effects of pixel-to-pixel errors and allows 
one to better remove hot pixels, bad columns, and charge traps from the image. Dithering also allows the recovery of information lost to undersampling by pixels that are not small 
compared to the point spread function (PSF). MultiDrizzle simplifies and automates the detection of cosmic-rays of these dithered observations. 
We used calibrated flat-fields from the HST pipeline and ran the MultiDrizzle script through the following steps. First, a static mask was created to 
identify bad pixels, then each image was sky-subtracted, shifts were determined from header coordinates for each image and were applied in drizzling each image separately 
onto registered output images. Next, a median image was created from these separate drizzled images and was blotted back to each original input image. 
Finally, the blotted images were used to compute cosmic ray masks, and the final drizzle combination was performed using these masks.

Prior to running MultiDrizzle 
the Planetary Camera (PC) data was removed from all images because its inclusion greatly increased the noise level of the output drizzled image. This was achieved by 
replacing the PC in each of the 24 images with hot pixels, forcing MultiDrizzle to automatically include the PC in the pixel mask for 
each individual image. We set the user-inputs to MultiDrizzle so the script would output separate science and weight images for these data that we have made
available online at: http://goods.gsfc.nasa.gov/release/UDF$_{-}$F300W.

\begin{figure*}
\epsscale{1}
\plotone{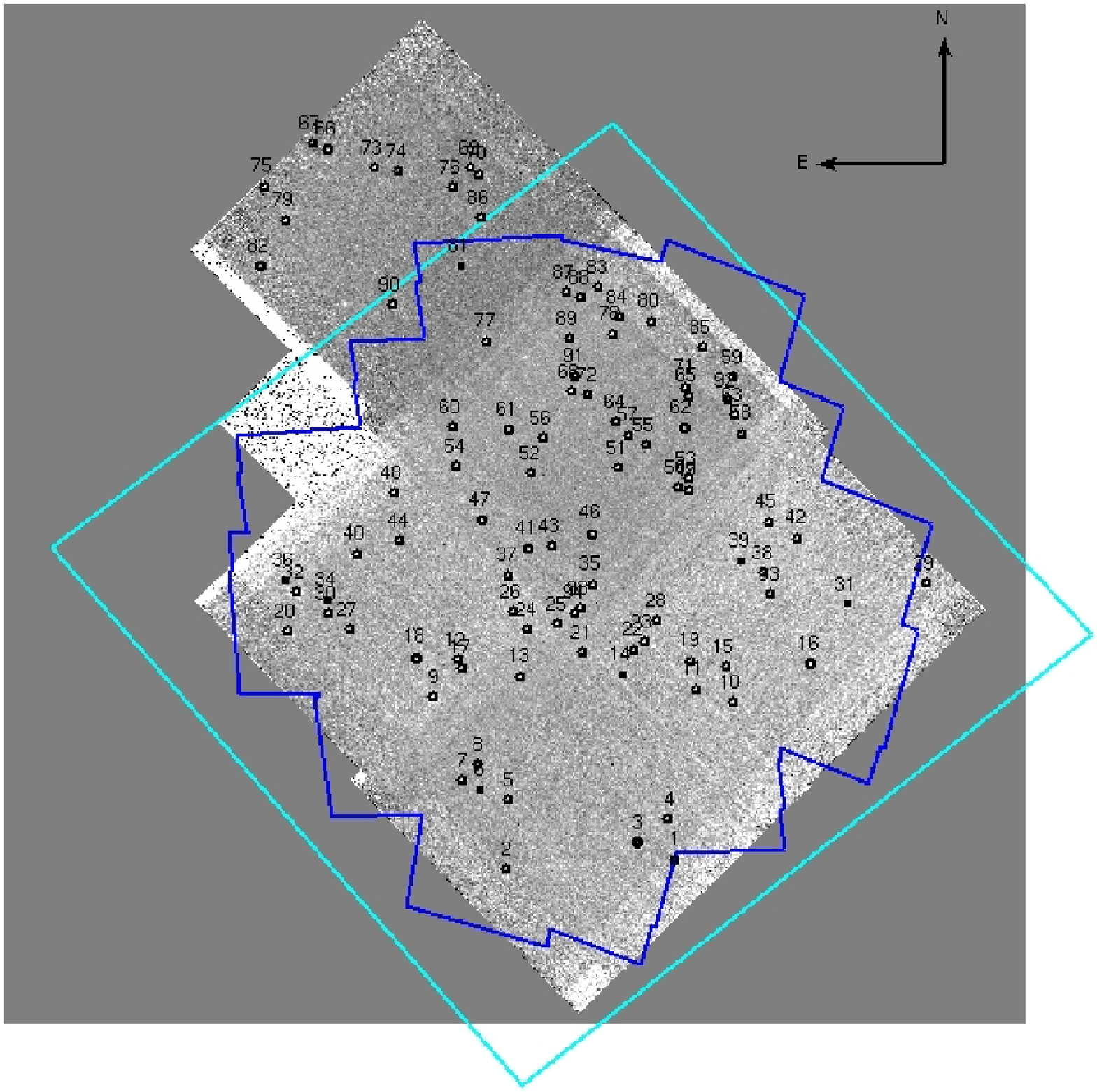}
\caption{WFPC2 drizzled U-band image overlayed with UDF(cyan) and ACS/SBC(blue) footprints. Note the large number of detections in the second rectangular region from the top of the image due to
a higher net exposure time for this region.}
\label{ubig}
\end{figure*}

The drizzled WFPC2 image shown in Figure \ref{ubig} has a total exposure time of 28,800s and a pixel scale of 0.1 arcsec. Due to the nature in which 
the 24 WFPC2 pointings were positioned, the final drizzled image does not have of a uniform depth. The majority of pointings overlap in the 
mid-upper region of the combined image. Consequently, the majority of UV sources are detected in this area of the UDF. Also, the upper-most section of the U-band 
image lies just outside the UDF footprint, and sources from this area are marked as such in the catalog. The reduction procedure used for the UDF ACS/SBC images 
is outlined in Teplitz et al. (2006).

\section{Object Identification And Catalog Preparation}

The catalog of U-band sources was produced using SourceExtractor version 2.5 (Bertin \& Arnouts 1996, hereafter SE). 
Initially, we created both a low-$\sigma$ (1.5$\sigma$ above background noise) and a high-$\sigma$ (3$\sigma$ above background noise) 
catalog. The high-$\sigma$ catalog contains all visually confirmed sources in the image, while the low-$\sigma$ catalog may contain
spurious detection. The difference between the SE parameters specified for these 
two catalogs was the detection threshold relative to the background RMS. For both catalogs, the minimum area of adjoining pixels 
for a detection was 15 pixels, and the minimum deblending parameter was set at 10\%, except for a few cases where we set the
deblending value in order to avoid multiple detections in one single object. 
This is particularly critical for source detection in U-band images since star-forming regions can appear 
as multiple clumps in one object. SE might detect those clumps as individual objects making false identifications. 

The GAIN, MAG$_{-}$ZEROPOINT, and SATUR$_{-}$LEVEL parameters which are specific to the WFPC2 camera were set to 7 e$^{-}$/ADU, 20.77 mag, and 2 ADU/s, respectively. 
The weight map produced by the drizzling process was used by setting the weight map type to MAP$_{-}$WEIGHT. 
Photometric measurements of each source were calculated using SE's automatic aperture magnitudes 
(MAG$_{-}$AUTO). MAG$_{-}$AUTO uses a Kron (1980) flexible elliptical aperture to measure the total magnitude 
of each source. Instead of using the classical aperture photomery with a fixed aperture radius, MAG$_{-}$AUTO has 
the advantage of limiting the background noise while detecting light from faint sources more effectively. 
The size of the background mesh which is 
subtracted from the photometry of each source (BACK$_{-}$SIZE) was set to 64 pixels, and its RMS value is used to calculate photometric errors. 
We cleaned the high-$\sigma$ and low-$\sigma$ catalogs removing sources with photometric errors $\ge$ 1.0. 

\begin{figure*}
\epsscale{1}
\plotone{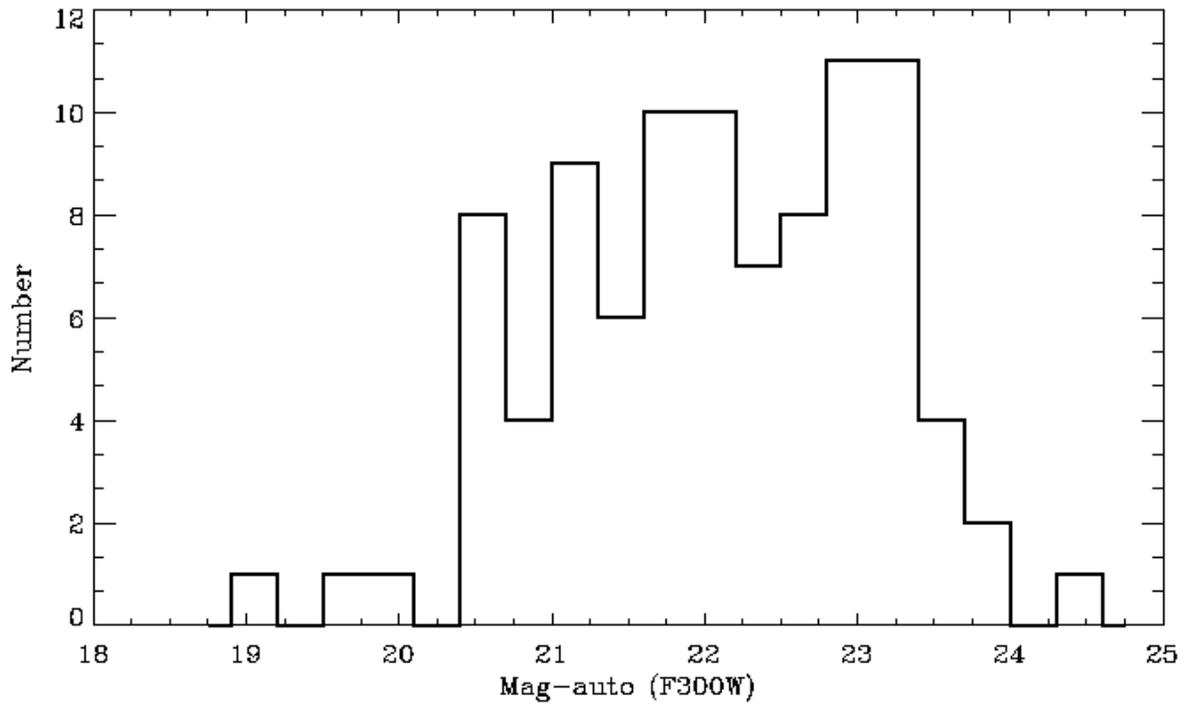}
\caption{Magnitude distribution of the U-band catalog, MAG$_{-}$AUTO from SE.} 
\label{mag}
\end{figure*}

The resulting U-band catalog includes all detections from the high-$\sigma$ catalog, and the remaining objects visually confirmed from 
the low-$\sigma$ catalog. The U-band 1.5$\sigma$ limiting magnitude measured within a 1$''$ diameter aperture is 23.5mag (AB) (Figure \ref{mag}). 

\subsection{Visual Identification Of U-Band Sources}

We have visually checked each SE U-band detection in order to decide (i) if a single source detection is actually multiple sources, (ii) 
if multiple source detections are single sources, (iii) if a detection is too noisy, or (iv) if there are any faint UV sources which are not detected. 
When such cases occur, SE parameters can be adjusted to maximize U-band source detections in the UDF image, and 
non-detections can be omitted from the catalog.  

We have also visually identified the U-band sources within a 
Hubble ACS/B-band image of the GOODS-S field that overlaps the UDF. The B-band sources had been cataloged by the GOODS team
using SE and have matched aperture photometry in multiple ACS bands (V,i,z) (Dahlen et al. 2007). 
The B-band catalog also lists spectral types (see \S 5 for definition) and photometric redshifts with a typical GOODS accuracy of $\Delta$$_{z}$=0.8 (where
$\Delta$$_{z}$$\equiv$$\langle$$|$z$_{phot}$-z$_{spec}$$|$/(1+z$_{spec}$)$\rangle$) (Dahlen et al. 2007) for each source. If a U-band detection could not be visually
identified as one of the objects in the B-band catalog it was removed from the U-band catalog. We did this because the B-band data is much deeper 
(limiting 10$\sigma$ sensativity is 27.8; 
Giavalisco et al. 2004) than the U-band, 
and we would not expect to detect a source in the U-band without also seeing it in the B-band. During this cleaning the majority of 
spurious U-band detections located in the borders of the WFPC2 image were removed. 
We also discovered five B-band objects that corresponded to multiple detections in the U-band. 
This was a result of setting SE's deblending parameters to a low value in an effort to detect as many U-band sources as possible. This parameter was adjusted 
in an additional SE run to obtain single detections of these sources. 
The final U-band catalog contains 96 objects.

\subsection{Catalog Matching}
We matched the final U-band catalog to the far-UV catalog of the UDF created from the ACS/SBC observations described in Section 2 (Siana et al. 2007). Each ACS/SBC
source was matched to the nearest U-band object within a 2.5$''$ radius.
The typical difference between WFPC2/F300W coordinates (RA and Dec) and ACS/SBC coordinates of the same object is 0.1$''$.
Thirteen U-band sources
do not have FUV detections because they are outside the ACS/SBC footprint. Any ACS/SBC sources with signal 
to noise $<$ 3$\sigma$ were not considered. In total, 31 of the 96 U-band objects have resolved matching ACS/SBC detections.

The UDF has also been observed with the Galaxy Evolution Explorer (GALEX) mission in the far and near ultraviolet 
(FUV: $\lambda$$_{\rm eff}$=1528\AA\ $\Delta$$\lambda$$_{FUV}$=269\AA\, 
NUV: $\lambda$$_{\rm eff}$=2271\AA\ $\Delta$$\lambda$$_{NUV}$=616\AA; GALEX field of view is 1$^{\circ}$.28 and 1$^{\circ}$.24 in FUV and NUV, 
and pixel scale is 1.5$''$/pixel) and is 
publicly available in the GALEX Release 4 (GR4) at the Multimission Archive at STScI (MAST). These data are from two different surveys, the All Sky Survey (AIS, 4$''$.3 FWHM) and the 
Deep Sky Survey (DIS, 5$''$.3 FWHM). The 5$\sigma$ limiting magnitudes of the AIS data are 20.8 in the NUV and 19.9 in the FUV, and for the DIS data 24.4 in the NUV and 24.8 in the FUV (Morrissey et al. 2007). 

\begin{figure*}
\epsscale{1.1}
\plotone{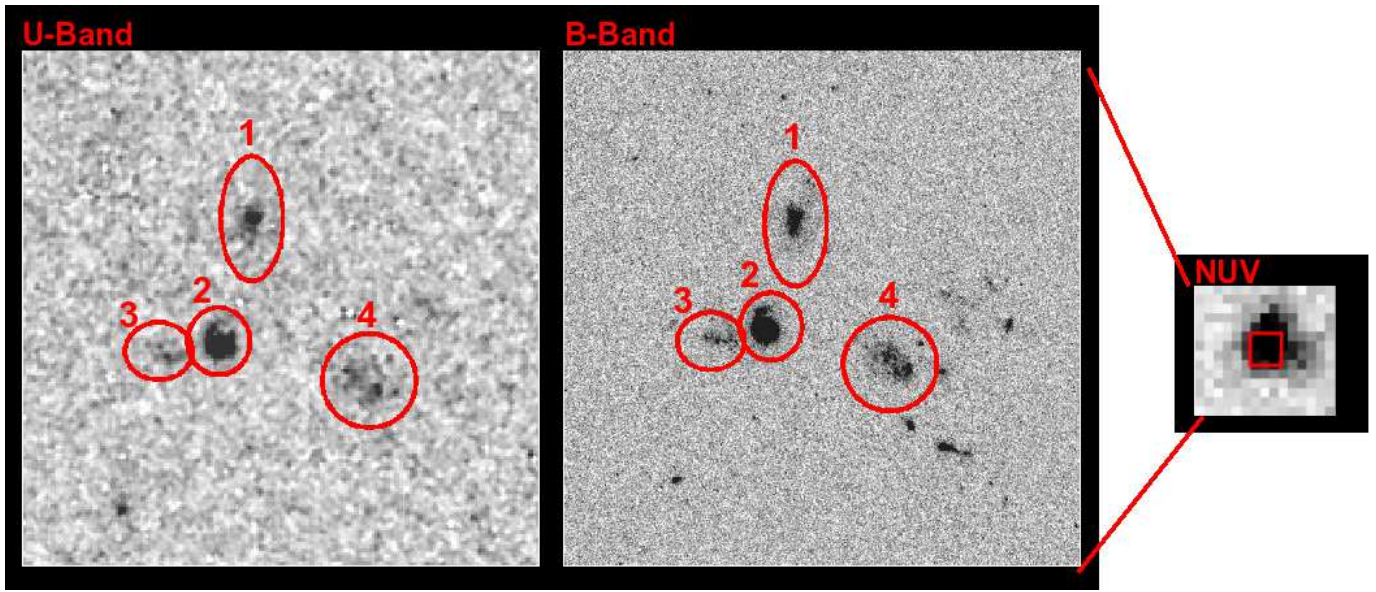}
\caption{WFPC2 (left), ACS (center), and GALEX (right) images showing four objects (1: 53.1619949 -27.7739410, 2: 53.1623802 -27.7750893, 3: below magnitude limit of the U-band catalog, 4: 
53.1608238 -27.7753963) which are within the GALEX beam. All three images are 20$''$ $\times$ 20$''$.}
\label{galex}
\end{figure*}

We searched for all U-band objects in the GR4 and found 28 detected by GALEX/FUV and 45 in the NUV. Since GALEX resolution is significantly lower than Hubble's (GOODS ACS image has 
0.03$''$/pixel) we have searched for objects where confusion might be problematic and one should use the 
GALEX data with caution. We found 22 objects where confusion was an important factor in the NUV and FUV and have flagged them in the catalog. In Figure \ref{galex} we show an example of 
a single GALEX NUV detection of at least four identified objects in the U and B-band. 

\section{The Catalog}   

\begin{figure*}
\epsscale{1}
\plotone{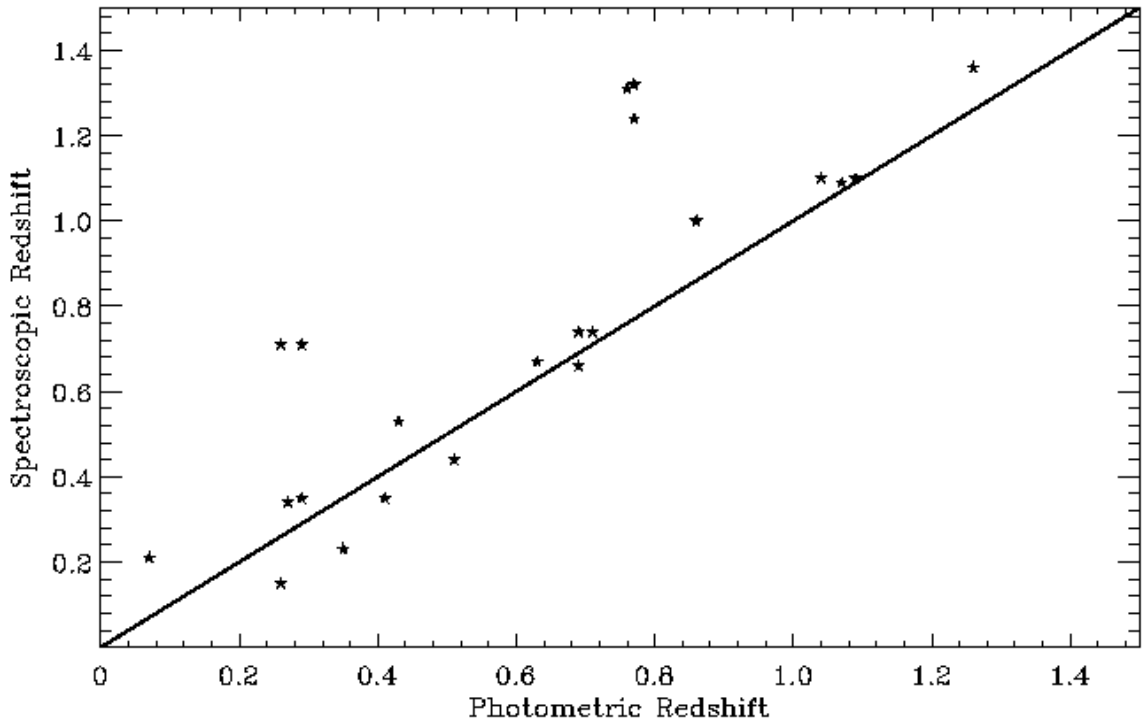}
\caption{Photometric Redshifts from GOODS vs. Spectroscopic Redshifts from the ESO/GOODS Chandra Deep Field-South spectroscopy masters catalog.}
\label{photspecz}
\end{figure*}

Table 1 presents the U-band catalog of the UDF NUV sources.  
Columns (1) and (2) are the GOODS World Coordinate System (WCS) Right Ascension and 
Declination in degrees. Three objects with Chandra X-ray detections 
(Koekemoer et al. 2004) are flagged as ``a" \footnote{Seven other Chandra X-ray sources found in the UDF were not detected in our U-band image.}. 
Columns (3)-(8) are the U, FUV, B, V, i, and z magnitudes and photometric errors (MAG$_{-}$AUTO), respectively. 
The B, V, i, and z magnitudes were obtained from the GOODS-S 
B-selected catalog. Note that not all U-band detections have FUV photometry because they are either outside the ACS/SBC 
footprint, are non-detections (S/N $<$ 3$\sigma$), or are not resolved 
in the FUV imaging. 
Columns (9) and (10) are GALEX NUV and FUV magnitudes from the GR4. Sources with confusion are flagged as ``c" in column 1.
Columns (11) and (12) list photometric redshifts (z$_{phot}$) for all objects, and available spectroscopic redshifts (z$_{spec}$) for 
21 objects. The z$_{spec}$ are from the GOODS collaboration (taken from the ESO/GOODS-S spectroscopy masters catalog\footnote[6]{See http://www.eso.org/science/goods/spectroscopy/CDFS$_{-}$Mastercat.}), 
and Figure \ref{photspecz} plots z$_{spec}$ as a function of z$_{phot}$ for these objects. Column (14) 
are spectral types from the GOODS B-selected catalog based on spectral energy distributions from Coleman et al. (1980) and Kinney et al. (1996). 
Type 1 galaxies are early-types (E, S0, Sa), type 2 are Sbc, type 3 are Scd, type 4 are irregulars, and types 5 and 6 are starbursts SB1 and SB2. 
The SED templates of SB1s and SB2s are differentiated by the values of intrinsic color excess, E(B-V). 
SB1 has E(B-V)$\le$0.10, and SB2 has 0.11$\le$E(B-V)$\le$0.21 (Kinney et al. 1996). 
The U-band catalog, including U, FUV, B, and BVi postage stamp images of each source, 
is available online at: http://goods.gsfc.nasa.gov/release/UDF$_{-}$F300W\\/original/gallery/udf$_{-}$u$_{-}$fuv.html.



\acknowledgments

We are grateful to the GOODS team. Support for this work was provided 
by NASA through grants HST-GO-10403.13A and HST-GO-10632.03A from the Space 
Telescope Science Institute, which is operated by the Association of 
Universities for Research in Astronomy, Inc., under NASA contract 
NAS5-26555. DFdM and E.V. were funded by NASA Research Grants NNG05GG06G and NNX08AR95H. 
We are grateful to Ms. Sara Petty for helping with the early stages of this project. 

GALEX is a NASA Small Explorer, launched in 2003 April. We gratefully 
acknowledge NASA's support for construction, operation, and science 
analysis for the GALEX mission, developed in cooperation with the 
Centre National d'Etudes Spatiales of France and the Korean Ministry 
of Science and Technology.

\clearpage

\clearpage

\begin{deluxetable}{lllcccclcclcll}
\tabletypesize{\scriptsize}
\rotate
\tablecaption{UDF U-band Sources}
\tablewidth{0pt}
\tablehead{
\colhead{RA} & 
\colhead{Dec} & 
\colhead{U$_{\rm mag}$} &
\colhead{FUV$_{\rm mag}$} &
\colhead{B$_{\rm mag}$} & 
\colhead{V$_{\rm mag}$} &
\colhead{i$_{\rm mag}$} & 
\colhead{z$_{\rm mag}$} & 
\colhead{GALEX} &
\colhead{GALEX} &
\colhead{z$_{\rm phot}$} & 
\colhead{z$_{\rm spec}$} & 
\colhead{ST} \\
\colhead{(Deg)} &
\colhead{(Deg)} &
\colhead{\null} &
\colhead{\null} &
\colhead{\null} &
\colhead{\null} &
\colhead{\null} &
\colhead{\null} &
\colhead{NUV$_{\rm mag}$} &
\colhead{FUV$_{\rm mag}$} &
\colhead{\null} &
\colhead{\null} &
\colhead{\null}
}
\tablecolumns{14}
\startdata
53.1306038&-27.7902641&20.50$\pm$0.05&       ...      &22.46$\pm$0.02&21.80$\pm$0.01&21.16$\pm$0.01&21.01$\pm$0.01&23.52$\pm$0.07&25.69$\pm$0.25&0.66&  ...   &3.67\\
53.1376457&-27.7919559&21.70$\pm$0.07&       ...      &23.54$\pm$0.04&23.13$\pm$0.02&22.44$\pm$0.03&22.04$\pm$0.02&	 ...     &	...	&0.85&  ...   &3.67\\
53.1409721&-27.7966537&22.12$\pm$0.05&24.97$\pm$0.06&23.54$\pm$0.04&22.47$\pm$0.01&21.95$\pm$0.02&21.69$\pm$0.02&24.63$\pm$0.24&	...	&0.41&0.3500&3.00\\
53.1421356&-27.7866974&21.02$\pm$0.03&23.55$\pm$0.03&22.11$\pm$0.01&21.15$\pm$0.01&20.73$\pm$0.01&20.57$\pm$0.01&23.18$\pm$0.09&23.90$\pm$0.16&0.35&0.2262&3.33\\
53.1445389&-27.7911339&21.58$\pm$0.06&       ...      &24.22$\pm$0.07&24.16$\pm$0.05&24.04$\pm$0.09&23.84$\pm$0.09&	 ...     &	...	&1.80&  ...   &6.00\\
53.1447334&-27.7854404&20.97$\pm$0.03&23.82$\pm$0.04&23.20$\pm$0.02&22.58$\pm$0.01&22.47$\pm$0.02&22.51$\pm$0.02&23.44$\pm$0.07&23.78$\pm$0.08&0.07&  ...   &6.00\\
53.1450996&-27.7894268&23.19$\pm$0.11&       ...      &24.15$\pm$0.06&23.96$\pm$0.04&23.58$\pm$0.06&23.07$\pm$0.05&	 ...     &      ...	  &0.84&  ...	&5.33\\
53.1470947\tablenotemark{c}&-27.7784290&22.46$\pm$0.08&       ...      &24.43$\pm$0.06&24.28$\pm$0.04&23.93$\pm$0.06&23.73$\pm$0.06&24.91$\pm$0.25&      ...	   &1.12&  ...   &6.00\\
53.1472054&-27.7884884&22.71$\pm$0.08&26.29$\pm$0.16&24.52$\pm$0.06&23.99$\pm$0.03&23.35$\pm$0.04&23.19$\pm$0.04&	...      &      ...	  &0.73&  ...	&4.00\\
53.1477814&-27.7769451&21.87$\pm$0.07&       ...      &24.74$\pm$0.06&24.66$\pm$0.05&23.90$\pm$0.05&23.43$\pm$0.04&	...      &      ...	  &1.07&1.0860&3.67\\
53.1478577\tablenotemark{c}&-27.7740345&20.58$\pm$0.03&       ...      &22.59$\pm$0.02&22.33$\pm$0.01&21.76$\pm$0.02&21.30$\pm$0.01&23.03$\pm$0.07&      ...	   &1.02&  ...   &4.00\\
53.1479225&-27.7996845&23.00$\pm$0.10&26.13$\pm$0.09&25.06$\pm$0.15&24.35$\pm$0.07&23.75$\pm$0.08&23.58$\pm$0.08&	...      &      ...	  &0.67&  ...	&4.00\\
53.1484451&-27.7757874&22.80$\pm$0.07&25.56$\pm$0.09&24.14$\pm$0.04&23.24$\pm$0.02&22.68$\pm$0.02&22.45$\pm$0.02&22.75$\pm$0.10&      ...       &0.48&3.6664&3.33\\
53.1485405&-27.7968845&22.48$\pm$0.08&       ...      &23.56$\pm$0.03&23.52$\pm$0.03&23.49$\pm$0.05&23.43$\pm$0.05&	...      &      ...       &1.84&  ...	&6.00\\
53.1506119&-27.7715912&20.51$\pm$0.04&22.73$\pm$0.02&22.24$\pm$0.02&21.54$\pm$0.01&21.26$\pm$0.01&21.19$\pm$0.01&22.62$\pm$0.07&22.93$\pm$0.09&0.23&  ...   &4.67\\
53.1512375&-27.7986679&21.51$\pm$0.06&25.86$\pm$0.09&24.93$\pm$0.08&24.37$\pm$0.04&23.64$\pm$0.04&23.83$\pm$0.06&	...      &      ...       &0.47&  ...	&6.00\\
53.1516838&-27.7964039&22.93$\pm$0.11&       ...      &23.77$\pm$0.04&23.49$\pm$0.03&23.35$\pm$0.05&23.15$\pm$0.05&	...      &      ...       &1.56&  ...	&5.67\\
53.1518784\tablenotemark{c}&-27.7754364&21.97$\pm$0.05&       ...      &23.76$\pm$0.05&23.36$\pm$0.03&22.58$\pm$0.03&22.17$\pm$0.02&22.93$\pm$0.06&      ...	   &1.00&  ...   &3.33\\
53.1518898\tablenotemark{c}&-27.7828751&22.73$\pm$0.04&       ...      &24.44$\pm$0.04&24.14$\pm$0.03&23.45$\pm$0.03&23.23$\pm$0.03&24.11$\pm$0.13&      ...	   &0.71&  ...   &4.00\\
53.1518974\tablenotemark{c}&-27.7819862&23.17$\pm$0.08&       ...      &24.60$\pm$0.06&24.32$\pm$0.04&23.72$\pm$0.05&23.32$\pm$0.04&24.19$\pm$0.18&      ...	   &0.84&  ...   &4.00\\
53.1520653\tablenotemark{c}&-27.7747822&21.65$\pm$0.03&       ...      &22.64$\pm$0.02&22.15$\pm$0.01&21.37$\pm$0.01&21.16$\pm$0.01&22.93$\pm$0.06&      ...	   &0.70&  ...   &3.67\\
53.1523628\tablenotemark{c}&-27.7779751&22.99$\pm$0.06&       ...      &24.71$\pm$0.08&24.84$\pm$0.07&24.96$\pm$0.16&24.29$\pm$0.10&22.93$\pm$0.06&      ...	   &1.55&  ...   &6.00\\
53.1528244\tablenotemark{c}&-27.7826958&23.57$\pm$0.06&28.01$\pm$0.26&25.52$\pm$0.10&25.21$\pm$0.07&24.61$\pm$0.09&24.25$\pm$0.08&24.11$\pm$0.13&      ...	 &0.70&  ...   &6.00\\
53.1531296&-27.8120804&21.04$\pm$0.04&       ...      &23.85$\pm$0.04&23.38$\pm$0.02&23.30$\pm$0.04&23.33$\pm$0.05&24.11$\pm$0.12&24.21$\pm$0.10&0.07&0.2128&6.00\\  
53.1536751&-27.8089409&22.06$\pm$0.06&25.89$\pm$0.14&24.92$\pm$0.08&24.12$\pm$0.03&23.80$\pm$0.05&23.62$\pm$0.05&	...      &      ...       &0.29&  ...	&5.33\\
53.1546783&-27.7932301&21.29$\pm$0.04&23.86$\pm$0.03&23.44$\pm$0.02&22.86$\pm$0.01&22.65$\pm$0.02&22.62$\pm$0.02&23.18$\pm$0.09&23.79$\pm$0.09&0.41&  ...   &6.00\\
53.1552696&-27.7695465&21.67$\pm$0.06&       ...      &23.65$\pm$0.05&23.07$\pm$0.02&22.38$\pm$0.03&22.13$\pm$0.03&	...      &      ...       &0.71&0.7359&3.67\\
53.1556816&-27.7793083&23.96$\pm$0.09&       ...      &23.99$\pm$0.05&23.68$\pm$0.03&23.28$\pm$0.04&23.16$\pm$0.04&	...     &      ...       &1.58&  ...   &4.00\\
53.1559105\tablenotemark{c}&-27.7948895&21.82$\pm$0.08&       ...      &24.02$\pm$0.04&23.79$\pm$0.03&23.43$\pm$0.05&23.11$\pm$0.04&24.63$\pm$0.12&      ...	   &0.91&  ...   &5.67\\
53.1564293&-27.8107758&20.45$\pm$0.02&24.69$\pm$0.09&22.20$\pm$0.02&21.62$\pm$0.01&20.97$\pm$0.01&20.84$\pm$0.01&22.90$\pm$0.08&      ...       &0.63&0.6650&3.67\\
53.1567726\tablenotemark{c}&-27.7955532&22.17$\pm$0.07&       ...      &23.79$\pm$0.04&23.31$\pm$0.02&22.66$\pm$0.03&22.08$\pm$0.02&23.34$\pm$0.15&      ...	   &1.09&1.0970&3.33\\
53.1572227&-27.7785568&22.78$\pm$0.06&       ...      &24.59$\pm$0.05&24.53$\pm$0.04&24.41$\pm$0.07&23.83$\pm$0.05&	...      &      ...       &0.76&1.3070&6.00\\
53.1578369&-27.7974815&22.27$\pm$0.10&       ...      &23.62$\pm$0.03&23.26$\pm$0.02&22.61$\pm$0.03&22.40$\pm$0.03&23.34$\pm$0.15&      ...       &0.70&  ...	&4.00\\
53.1580658&-27.7692299&21.92$\pm$0.02&       ...      &21.18$\pm$0.01&20.40$\pm$0.00&20.02$\pm$0.01&19.91$\pm$0.00&23.79$\pm$0.14&      ...       &0.11&  ...	&3.00\\
53.1581879&-27.7811279&21.00$\pm$0.02&24.10$\pm$0.04&22.93$\pm$0.02&22.43$\pm$0.01&21.93$\pm$0.02&21.90$\pm$0.02&23.22$\pm$0.09&25.13$\pm$0.34&0.44&  ...   &5.67\\
53.1583176&-27.7774792&22.76$\pm$0.04&       ...      &24.77$\pm$0.06&24.79$\pm$0.05&24.75$\pm$0.09&24.64$\pm$0.10&	...      &      ...       &1.88&  ...	&6.00\\
53.1587791&-27.7705669&22.47$\pm$0.04&       ...      &23.90$\pm$0.04&23.56$\pm$0.03&22.84$\pm$0.03&22.45$\pm$0.02&23.86$\pm$0.22&      ...       &0.91&  ...	&3.67\\
53.1599007&-27.7668762&23.39$\pm$0.27&       ...      &24.76$\pm$0.07&24.72$\pm$0.06&24.02$\pm$0.07&23.59$\pm$0.06&21.86$\pm$0.04&      ...       &0.90&  ...	&4.00\\
53.1604424&-27.7903595&23.03$\pm$0.07&       ...      &24.01$\pm$0.04&24.01$\pm$0.03&23.92$\pm$0.05&23.92$\pm$0.06&	...      &      ...       &1.82&  ...	&6.00\\
53.1604996&-27.7863064&23.83$\pm$0.07&       ...      &25.50$\pm$0.15&25.36$\pm$0.11&24.55$\pm$0.11&24.28$\pm$0.10&	...      &      ...       &0.90&  ...	&4.00\\
53.1608238&-27.7753963&21.70$\pm$0.06&       ...      &23.44$\pm$0.04&22.50$\pm$0.02&21.67$\pm$0.02&21.42$\pm$0.02&22.44$\pm$0.06&23.02$\pm$0.10&0.64&  ...   &3.00\\
53.1613579&-27.7957401&22.93$\pm$0.10&       ...      &24.48$\pm$0.06&24.06$\pm$0.04&23.77$\pm$0.06&23.54$\pm$0.06&24.62$\pm$0.25&25.37$\pm$0.24&1.00&  ...   &5.67\\
53.1615295\tablenotemark{c}&-27.7676716&22.95$\pm$0.10&26.58$\pm$0.15&24.96$\pm$0.07&24.45$\pm$0.04&23.78$\pm$0.05&23.70$\pm$0.05&21.86$\pm$0.00&      ...	 &0.69&  ...   &4.00\\
53.1615944\tablenotemark{a}&-27.7922535&20.89$\pm$0.02&22.65$\pm$0.02&21.81$\pm$0.01&21.37$\pm$0.01&21.02$\pm$0.01&20.77$\pm$0.01&21.98$\pm$0.04&23.08$\pm$0.09&0.42&  ...   &5.33\\
53.1619873\tablenotemark{c}&-27.7925415&21.63$\pm$0.04&24.74$\pm$0.05&24.91$\pm$0.04&24.16$\pm$0.02&23.75$\pm$0.03&23.71$\pm$0.03&21.97$\pm$0.03&22.95$\pm$0.04&0.25&  ...   &5.00\\
53.1619949\tablenotemark{c}&-27.7739410&21.87$\pm$0.06&23.95$\pm$0.04&23.58$\pm$0.03&22.83$\pm$0.02&22.62$\pm$0.02&22.53$\pm$0.03&22.39$\pm$0.04&23.12$\pm$0.05&0.26&  ...   &5.67\\
53.1623802\tablenotemark{c}&-27.7750893&21.16$\pm$0.03&23.66$\pm$0.03&22.45$\pm$0.02&21.48$\pm$0.01&20.97$\pm$0.01&20.69$\pm$0.01&22.39$\pm$0.04&23.12$\pm$0.05&0.37&  ...   &3.00\\
53.1624947&-27.7709045&24.45$\pm$0.15&       ...      &25.46$\pm$0.11&25.10$\pm$0.07&24.61$\pm$0.09&24.45$\pm$0.09&23.86$\pm$0.22&      ...       &1.13&  ...	&6.00\\
53.1628456\tablenotemark{ac}&-27.7672405&21.16$\pm$0.03&    ...	   &21.17$\pm$0.01&20.94$\pm$0.01&20.87$\pm$0.01&20.84$\pm$0.01  &21.96$\pm$0.03&      ...	 &0.05&  ...   &6.00\\
53.1635971&-27.7935085&21.89$\pm$0.07&       ...      &25.17$\pm$0.07&25.09$\pm$0.05&24.98$\pm$0.08&24.47$\pm$0.07&	...      &      ...       &1.54&  ...	&6.00\\
53.1641121&-27.7873249&23.08$\pm$0.05&       ...      &24.95$\pm$0.07&24.78$\pm$0.05&24.32$\pm$0.07&24.00$\pm$0.06&	...      &      ...       &0.75&  ...	&6.00\\
53.1648941&-27.7787838&23.27$\pm$0.12&       ...      &25.31$\pm$0.10&25.34$\pm$0.08&24.89$\pm$0.11&24.63$\pm$0.11&	...      &      ...       &1.26&  ...	&6.00\\
53.1659012&-27.7815647&22.96$\pm$0.05&       ...      &24.67$\pm$0.08&24.05$\pm$0.04&23.29$\pm$0.04&22.79$\pm$0.03&	...      &      ...       &0.95&  ...	&3.33\\
53.1661797&-27.7875214&21.70$\pm$0.05&       ...      &23.02$\pm$0.03&22.64$\pm$0.02&21.98$\pm$0.02&21.41$\pm$0.01&	...      &      ...       &1.04&1.0951&3.33\\
53.1662140&-27.7939320&21.99$\pm$0.10&       ...      &24.56$\pm$0.06&24.54$\pm$0.04&23.99$\pm$0.05&23.63$\pm$0.04&23.48$\pm$0.24&      ...       &0.71&  ...	&6.00\\
53.1668854&-27.7976780&22.72$\pm$0.06&       ...      &24.37$\pm$0.05&24.43$\pm$0.04&23.93$\pm$0.05&23.75$\pm$0.05&24.39$\pm$0.23&      ...       &1.18&  ...	&6.00\\
53.1675873\tablenotemark{c}&-27.7925072&23.27$\pm$0.54&       ...      &23.76$\pm$0.04&23.45$\pm$0.02&23.01$\pm$0.03&22.72$\pm$0.03&23.48$\pm$0.24&      ...	   &0.71&  ...   &5.67\\
53.1679382&-27.7781277&23.47$\pm$0.12&26.16$\pm$0.10&24.86$\pm$0.08&24.05$\pm$0.03&23.70$\pm$0.05&23.63$\pm$0.06&	...      &      ...       &0.26&0.7122&5.33\\
53.1680222&-27.7896690&22.09$\pm$0.07&25.64$\pm$0.09&24.12$\pm$0.04&23.57$\pm$0.02&22.98$\pm$0.03&22.90$\pm$0.03&24.86$\pm$0.14&&0.50&  ...   &4.33\\
53.1680603\tablenotemark{c}&-27.8074017&22.91$\pm$0.11&25.32$\pm$0.08&24.77$\pm$0.09&23.97$\pm$0.04&23.72$\pm$0.06&23.70$\pm$0.07&24.35$\pm$0.17&25.89$\pm$0.27&0.41&  ...   &6.00\\
53.1681747&-27.8128967&23.16$\pm$0.07&       ...      &24.43$\pm$0.06&24.06$\pm$0.03&23.35$\pm$0.04&23.04$\pm$0.03&	...      &      ...       &0.83&  ...	&3.67\\
53.1699409&-27.7710609&20.59$\pm$0.02&       ...      &22.01$\pm$0.02&21.18$\pm$0.01&20.37$\pm$0.01&20.12$\pm$0.01&22.94$\pm$0.05&25.26$\pm$0.20&0.65&  ...   &3.00\\
53.1703148&-27.7852764&23.07$\pm$0.09&25.32$\pm$0.07&24.86$\pm$0.07&24.15$\pm$0.03&23.90$\pm$0.04&23.67$\pm$0.04&	...      &      ...       &0.29&0.7122&5.67\\
53.1704826\tablenotemark{b}&-27.7613792&21.58$\pm$0.03&       ...      &22.53$\pm$0.01&21.84$\pm$0.01&21.49$\pm$0.01&21.39$\pm$0.01&23.16$\pm$0.09&23.91$\pm$0.09&0.25&  ...   &4.67\\
53.1705437&-27.8065834&22.75$\pm$0.09&       ...      &24.65$\pm$0.06&24.67$\pm$0.05&24.42$\pm$0.07&23.96$\pm$0.06&24.74$\pm$0.47&      ...       &0.77&1.2441&6.00\\
53.1706848\tablenotemark{bc}&-27.7579365&22.42$\pm$0.05&      ...      &24.19$\pm$0.08&23.56$\pm$0.03&22.98$\pm$0.03&22.93$\pm$0.04&22.77$\pm$0.06&23.16$\pm$0.10&0.27&  ...   &5.00\\
53.1707726&-27.8046780&23.14$\pm$0.09&       ...      &24.06$\pm$0.05&23.91$\pm$0.04&23.71$\pm$0.06&23.63$\pm$0.07&25.04$\pm$0.24&      ...       &1.79&  ...	&6.00\\
53.1713791\tablenotemark{bc}&-27.7574749&20.96$\pm$0.02&      ...      &21.69$\pm$0.02&21.07$\pm$0.01&20.71$\pm$0.01&20.63$\pm$0.01&22.70$\pm$0.04&23.23$\pm$0.05&0.25&  ...   &5.33\\
53.1721077\tablenotemark{c}&-27.7969379&23.11$\pm$0.13&25.21$\pm$0.06&24.58$\pm$0.08&23.71$\pm$0.03&23.42$\pm$0.05&23.32$\pm$0.05&22.94$\pm$0.05&23.50$\pm$0.06&0.19&  ...   &3.33\\
53.1721840&-27.8058681&22.21$\pm$0.06&       ...      &23.94$\pm$0.06&23.83$\pm$0.04&23.71$\pm$0.08&23.24$\pm$0.06&	...      &      ...       &0.77&1.3180&6.00\\
53.1722603&-27.7651482&22.41$\pm$0.08&       ...      &23.34$\pm$0.04&22.64$\pm$0.02&22.01$\pm$0.02&21.88$\pm$0.02&23.98$\pm$0.16&24.65$\pm$0.25&0.51&  ...   &3.67\\
53.1725159\tablenotemark{c}&-27.7963371&21.05$\pm$0.06&23.66$\pm$0.03&23.01$\pm$0.03&22.15$\pm$0.01&21.85$\pm$0.02&21.67$\pm$0.02&22.89$\pm$0.08&23.27$\pm$0.10&0.29&0.3469&3.67\\
53.1726112&-27.7809887&23.37$\pm$0.15&       ...      &24.04$\pm$0.06&22.90$\pm$0.02&22.05$\pm$0.02&21.71$\pm$0.01&	...      &      ...       &0.66&  ...	&2.67\\
53.1730003&-27.7779026&22.08$\pm$0.06&       ...      &24.06$\pm$0.04&23.98$\pm$0.03&23.53$\pm$0.04&23.20$\pm$0.04&24.46$\pm$0.26&      ...       &0.71&  ...	&6.00\\
53.1730042\tablenotemark{bc}&-27.7590351&20.62$\pm$0.02&      ...      &21.51$\pm$0.01&20.76$\pm$0.01&20.37$\pm$0.01&20.26$\pm$0.01&22.69$\pm$0.06&22.99$\pm$0.08&0.26&  ...   &5.00\\
53.1747513&-27.7992420&19.58$\pm$0.03&22.00$\pm$0.02&21.14$\pm$0.01&20.54$\pm$0.01&20.25$\pm$0.01&20.14$\pm$0.01&22.04$\pm$0.03&22.30$\pm$0.03&0.26&0.1514&5.67\\
53.1761894&-27.7961178&21.25$\pm$0.06&       ...      &22.87$\pm$0.02&22.36$\pm$0.01&21.62$\pm$0.01&21.23$\pm$0.01&24.28$\pm$0.18&      ...       &0.86&0.9961&3.67\\
53.1767311\tablenotemark{d}&-27.7996502&22.89$\pm$0.12&       ...      &22.53$\pm$0.02&20.58$\pm$0.00&18.93$\pm$0.00&18.32$\pm$0.00&  ...  &  ...  &0.70&  ...   &1.33\\
53.1777191&-27.7869625&22.52$\pm$0.07&       ...      &25.15$\pm$0.08&25.08$\pm$0.07&24.52$\pm$0.08&24.41$\pm$0.08&	...      &      ...       &1.12&  ...	&6.00\\
53.1778793\tablenotemark{b}&-27.7577496&23.50$\pm$0.12&       ...      &25.02$\pm$0.05&25.05$\pm$0.05&24.44$\pm$0.06&24.27$\pm$0.06&	 ...	&      ...	 &0.91&  ...   &6.00\\
53.1782227&-27.7830944&21.45$\pm$0.05&       ...      &24.03$\pm$0.05&23.84$\pm$0.03&23.45$\pm$0.04&22.92$\pm$0.03&	...      &      ...       &0.77&  ...	&5.67\\  
53.1784172&-27.7682304&21.09$\pm$0.05&       ...      &22.39$\pm$0.02&21.54$\pm$0.01&20.68$\pm$0.01&20.39$\pm$0.01&23.12$\pm$0.07&      ...       &0.72&  ...	&3.00\\
53.1799660\tablenotemark{b}&-27.7573910&23.55$\pm$0.13&       ...      &25.28$\pm$0.08&24.76$\pm$0.05&24.31$\pm$0.06&24.33$\pm$0.08&	 ...	&      ...	 &0.68&  ...   &6.00\\
53.1815453&-27.7879925&22.07$\pm$0.04&23.86$\pm$0.03&23.94$\pm$0.04&23.48$\pm$0.02&23.35$\pm$0.03&23.44$\pm$0.04&23.57$\pm$0.14&23.55$\pm$0.19&0.07&0.2122&6.00\\
53.1821976&-27.7939968&23.18$\pm$0.09&27.64$\pm$0.23&25.70$\pm$0.09&25.59$\pm$0.06&25.07$\pm$0.08&24.93$\pm$0.08&	...      &      ...       &1.13&  ...	&6.00\\
53.1826210\tablenotemark{d}&-27.7681408&19.44$\pm$0.00&       ...      &       ...      &	    ...	   &	   ...	  &17.45$\pm$0.00&  ...  &  ...  &0.30&  ...   &6.00\\
53.1841164\tablenotemark{b}&-27.7559299&22.82$\pm$0.13&       ...      &24.89$\pm$0.06&24.73$\pm$0.05&24.41$\pm$0.07&24.33$\pm$0.08&	 ...	&      ...	 &1.89&  ...   &6.00\\
53.1841660&-27.7926407&21.36$\pm$0.02&25.92$\pm$0.13&22.90$\pm$0.02&22.33$\pm$0.01&21.65$\pm$0.01&21.44$\pm$0.01&    	  ...	 &	...	  &0.69&0.7372&3.67\\
53.1841698&-27.7915535&23.38$\pm$0.10&       ...      &24.95$\pm$0.14&24.48$\pm$0.07&23.77$\pm$0.07&23.36$\pm$0.06&    	  ...	 &	...	&1.45&  ...   &3.00\\
53.1855202\tablenotemark{b}&-27.7554283&22.01$\pm$0.07&      ...      &23.96$\pm$0.03&23.07$\pm$0.01&22.52$\pm$0.02&22.36$\pm$0.02&	 ...	&      ...	 &0.43&0.5328&3.33\\
53.1869583\tablenotemark{c}&-27.7910004&19.03$\pm$0.01&21.67$\pm$0.01&20.21$\pm$0.01&19.19$\pm$0.00&18.65$\pm$0.00&18.44$\pm$0.00&21.32$\pm$0.02&21.89$\pm$0.02&0.23&  ...   &2.33\\
53.1877899\tablenotemark{c}&-27.7940979&19.83$\pm$0.02&22.23$\pm$0.02&21.49$\pm$0.01&20.68$\pm$0.01&20.45$\pm$0.01&20.23$\pm$0.01&21.95$\pm$0.02&22.49$\pm$0.03&0.27&0.3446&3.67\\
53.1879463\tablenotemark{b}&-27.7615261&21.55$\pm$0.06&       ...      &21.50$\pm$0.01&20.61$\pm$0.01&20.22$\pm$0.01&20.05$\pm$0.01&23.59$\pm$0.11&24.47$\pm$0.21&0.22&  ...   &3.33\\
53.1879768\tablenotemark{ac}&-27.7900066&21.10$\pm$0.03&24.86$\pm$0.06&22.69$\pm$0.02&21.53$\pm$0.01&20.79$\pm$0.01&20.39$\pm$0.01&21.32$\pm$0.02&21.89$\pm$0.02&0.51&0.4357&2.33\\
53.1898384\tablenotemark{b}&-27.7588539&20.64$\pm$0.03&       ...      &21.84$\pm$0.01&21.41$\pm$0.01&21.17$\pm$0.01&21.17$\pm$0.01&22.43$\pm$0.04&22.65$\pm$0.03&0.24&  ...   &6.00\\
53.1901550\tablenotemark{b}&-27.7651958&20.49$\pm$0.06&       ...      &22.01$\pm$0.02&20.95$\pm$0.01&20.52$\pm$0.01&20.29$\pm$0.01&22.82$\pm$0.07&23.69$\pm$0.13&0.36&  ...   &3.00\\
\enddata        

\tablenotetext{a}{X-ray source (Koekemoer et al. 2004)}
\tablenotetext{b}{Source is outside the UDF footprint}
\tablenotetext{c}{Confusion in GALEX image}
\tablenotetext{d}{Star}
                                                                                                                                                  
\end{deluxetable}

\end{document}